# Exploratory spatial data analysis with gwpcorMapper: an interactive mapping tool for geographically weighted correlation and partial correlation

J. E. H. Percival[1], N Tsutsumida[2,*], D Murakami[3], T Yoshida[4] and T Nakaya[5]

[1] Graduate School of Agriculture, Kyoto University

[2*] Graduate School of Science and Engineering, Saitama University

[3] Department of Statistical Modeling, The Institute of Statistical Mathematics

[4] Department of Urban Engineering, The University of Tokyo

[5] Graduate School of Environmental Studies, Tohoku University

*Correspondence author

**Abstract**
Exploratory spatial data analysis (ESDA) plays a key role in research that includes geographic data. In ESDA, analysts typically need to visualize observations and local relationships on a map. However, software dedicated to visualizing local degrees of association between multiple variables in high-dimensional geospatial datasets remains undeveloped. This paper introduces *gwpcorMapper*, a newly developed software application for mapping geographically weighted (GW) correlation and partial correlation statistics in high-dimensional datasets. *gwpcorMapper* facilitates ESDA by giving researchers the ability to interact with map components that describe local correlative relationships. *gwpcorMapper* is open source and is built using the R Shiny framework. The software inherits its algorithm logic from *GWpcor*, an R library for calculating the geographically weighted correlation and partial correlation statistics. We demonstrate the application of *gwpcorMapper* by using it to investigate census data to find meaningful relationships that describe the work-life environment in the 23 special wards of Tokyo, Japan. *gwpcorMapper* is useful in both variable selection and parameter tuning for GW statistics.

## 1. Introduction

In exploratory spatial data analysis (ESDA), researchers investigate patterns and processes in data using statistical methods that explicitly take the spatial context of the data into consideration. It is often considered an essential first step in any geospatial study as it is during ESDA that researchers may discover previously unknown associations and form hypotheses (Fotheringham 1992; Haining, Wise, and Ma 1998; Unwin and Unwin 1998). The general procedure by which ESDA occurs consists of steps of collecting and selecting data according to the target topic, exploring spatial properties of data by analytical methods through an iterative process of changing variables and parameter values, and visualizing data using dynamic and interactive graphics in a Geographic Information System (GIS) (Brunsdon 1998; Dykes 1998; Anselin 1999; Anselin 2005; Dykes and Brunsdon 2007; Bivand 2010).

1.1 Spatial Dependence, Spatial Heterogeneity, and the Evolution of ESDA

Central to ESDA is examining if the data exhibit evidence of underlying properties of spatial processes. The most common of which are spatial dependence and spatial heterogeneity (Anselin and Rey 1991; Getis 1994). Spatial dependence has long been a major focus in ESDA (Cliff and Ord 1973; Anselin 1988a; Anselin and Rey 1991). It characterizes how observations of a variable may depend on both the location and the relative geographic distance to other observations of the same variable, suggesting that data that are spatially proximate are more closely related (Anselin 1988a; Getis 1994). Spatial heterogeneity, on the other hand, describes the variability of observations of or between data over space and can thus highlight any spatially non-stationary processes (Getis 1994; Anselin 1988b). It has been noted that of the two, spatial heterogeneity has received less attention in the context of ESDA where analytical methods typically focus on identifying degrees of spatial autocorrelation (Lee 2009; Comber et al. 2022). Yet observing spatial heterogeneity, especially through localized approaches, remains an essential form of spatial data analysis. Summarizing data with descriptive statistics that encompass the entirety of the dataset without any special accommodation for locality, such as a global mean or standard deviation, can be an initial step of ESDA but such simple global analyses can lead to undesirably high levels of descriptive data loss as such methods fail completely in detecting spatially varying relationships.

Geographically weighted (GW) statistics have been introduced to address the inability of global descriptive statistics to capture locally varying relationships across multiple variables (Brunsdon, Fotheringham, and Charlton 1996). These encompass a suite of localized statistical methods that have been developed for observing and handling spatial heterogeneity in data by generating statistics under a distance-decay weighted moving window that is traversed across the spatial surface of the data (Lu et al. 2014; Gollini et al. 2015). GW regression (GWR) (Brunsdon, Fotheringham, and Charlton 1996; Fotheringham, Charlton, and Brunsdon 1997), for example, has emerged as one of the most widely applied methods for modeling predictors of spatially varying relationships (Comber et al. 2022). In GWR, the relationship between the response variable and a set of spatially weighted predicators is observed at each location in space and mapped (Brunsdon, Fotheringham, and Charlton 1996). Since the introduction of GWR, other GW-based statistical methods have been proposed, including GW summary statistics (including mean, variance, and correlation) (Brunsdon, Fotheringham, and Charlton 2002; Harris and Brunsdon 2010), GW principal components analysis (GWPCA) (Lloyd 2010; Harris, Brunsdon, and Charlton 2011), and GW discriminant analysis (GWDA) (Brunsdon, Fotheringham, and Charlton 2007).

Of the many GW statistical methods, GW summary statistics, particularly GW correlation (Brunsdon, Fotheringham, and Charlton 2002; Harris and Brunsdon 2010) and GW partial correlation (Percival and Tsutsumida 2017), stand out in terms of ESDA as these methods are very effective in understanding local correlations in multivariate datasets with underlying spatial heterogeneity. For example, Harris and Brunsdon (2010) demonstrated how exploring local correlation coefficients can be used to explain local drivers of freshwater acidification processes and create spatially driven models of critical load in freshwater systems. Their results show how GW correlation analyses highlight important local characteristics of data and help determine variable selection for more sophisticated successive analyses, such as GWPCA or GWR.

1.2 Limitations in Existing Software
Despite the significance of GW correlation in ESDA, existing software applications have focused primarily on visual inspection of map data or on representations of spatial dependence (Bivand 2010). For example, desktop software like *geoDa* (Anselin, Syabri, and Kho 2006), *ArcMap* (ESRI 2020), *SpaceStat* (BioMedware 2020), and *QGIS* (QGIS Development Team 2021) all have excellent features for a variety of map visualizations, including quantile maps, thematic maps, cartograms and choropleth maps; yet in terms of analytical methods in ESDA, the focus is primarily on highlighting degrees of spatial autocorrelation. None, to our knowledge, provide GW correlative analyses and are therefore limited in their ability to encompass local ESDA with geospatial datasets in a manner that explicitly considers spatial heterogeneity between multiple variables. Dykes and Brunsdon (2007) introduced visualization concepts for ESDA with GW statistics, including correlation, however their related software is not available. Instead, software libraries such as *GWmodel* (Lu et al. 2014; Gollini et al. 2015), *spgwr* (Bivand and Yu 2020), or *lctools* (Kalogirou 2020) in the R programming language have provided the means for calculating GW correlative statistics. However, as libraries they lack an interactive graphical user interface (GUI), and this can make visualizing spatial patterns from their supported functions across varying parameters and multiple variables in high-dimensional datasets overly complex (Dykes and Brunsdon 2007).

Furthermore, these libraries lack the ability to calculate GW partial correlation and can have poor computational performance with observation-rich data.

**Table 1.** A comparison between commonly used software that include tools for ESDA and the newly proposed software: gwpcorMapper.

| Software | Has GUI | GW Correlation | GW Partial Correlation |
|---|---|---|---|
| GWmodel | No | Yes | No |
| spgwr | No | Yes | No |
| lctools | No | Yes | No |
| GeoDa | Yes | No | No |
| QGIS | Yes | No | No |
| ArcMap | Yes | No | No |
| SpaceStat | Yes | No | No |
| gwpcorMapper | Yes | Yes | Yes |

1.3 Interactive Visualization and Parameter Selection in ESDA

Presenting and summarizing data in an interactive visual manner has been central to ESDA, as it is through such interaction that users can explore complex data structures and build knowledge (Thomas and Cook 2006; Dykes and Brunsdon 2007; Comber et al. 2022). Such becomes increasingly important with very large high-dimensional datasets as these graphics can help simplify the complexity of big spatial data and can guide decisions in model design (Comber et al. 2022). For example, it can be difficult to conceptualize how the addition a new variable or a change in model parameters may affect a model's output. Inspecting maps that explore these changes help researchers and analysts make sense of their models and understand central concepts like the effect of scale.

The ability to explore varying parameters in a GW model is particularly important for ESDA. Given that GW correlation and partial correlation operate under a moving window, parameters that control both the size and the shape of the moving window are crucial. Such parameters are often referred to as "bandwidth" and "kernel" in both the literature and existing software applications. Bandwidth controls for the effects of spatial scale on the analysis while kernel type defines the weighting scheme employed. Software libraries that support GW correlative statistics may also allow users to select either parametric or non-parametric calculations, such as Pearson's or Spearman's methods, when performing ESDA. Parameter selection and tuning are recognized as important yet difficult aspects of implementing GW summary statistics with existing software, as it is often recommended to select parameters values based on data interpretation (Fotheringham, Brunsdon, and Charlton 2003; Harris and Brunsdon 2010). For example, while a cross-validation (CV) scheme may be implemented to determine a statistically optimized bandwidth size for GWR (Farber and Páez, 2007), the CV approach will fail to yield an appropriate result for GW summary statistics as it investigates the fitness of the statistical model and thus does not function to serve GW summary statistics. Instead, an iterative process of adding and removing variables, changing kernel type and size, and correlation methods in an interactive manner may provide the setting for researchers to gain valuable insights of the data. Software that can aid in parameter selection when calculating GW summary statistics will provide important benefits to the process of ESDA.

1.4 Study Objectives

The objective of this study is to demonstrate an enhanced process for ESDA that overcomes the gap in existing software for local correlative analyses in high dimensional geospatial data. To achieve this, we developed a software application called *gwpcorMapper*. This software allows for ESDA using GW correlation and GW partial correlation analyses by enabling users to interactively select covariates and change bandwidth sizes, kernel types, and correlation methods. It complements existing tools by making GW correlation and GW partial correlation analyses simpler and supports the exploratory process by giving users the ability to investigate multiple variable and parameter choices in an interactive manner. *gwpcorMapper* contributes to the ESDA iterative process by enabling fast and simple inspection of the level of spatial heterogeneity between multiple variables and at various spatial scales to find meaningful local relationships.

We explain the application of *gwpcorMapper* and demonstrate its use in ESDA with census statistics of the 23 special wards in Tokyo, Japan for the year 2005 and 2006. This data consists of 228 variables that describe the urban social structure of Tokyo within the 3134 *chocho-aza* (the smallest administrative unit in Japan) of the 23 special wards. We investigated the data using *gwpcorMapper* to find meaningful covariates and parameter values of bandwidth and kernel type. We then used these results to focus on local characteristics of urban social structure to discuss the working and living environment in Tokyo in the context of the compact city paradigm, a popular urban planning concept that emphasizes high residential population density in cities with highly diversified land use (Dantzig and Saaty 1973; Bibri, Krogstie, and Kärrholm 2020). The diverse and tightly packed structure of compact cities encourage lower commuting times and supports both walking and more efficient public transportation systems making them often seen as a strong contributor to sustainability within cities. Through this case study, we demonstrate how *gwpcorMapper* can aid in the ESDA process. Finally, we provide computational benchmarks and a comparison with existing similar software that calculate GW correlation statistics to underline the validity of *gwpcorMaper* for ESDA in high dimensional geospatial data.

## 2. Software Implementation
### 2.1 Overview of *gwpcormapper*

*gwpcorMapper* is a web application that is built with R (R Core Team 2020) using the Shiny framework (Chang et al. 2020) and is published as open source under the General Public License v3.0. R is one of the most popular and actively used software environments for geospatial statistical analysis and R shiny provides the means for turning R programs into interactive web applications. As an R Shiny application, *gwpcorMapper* can be launched simply in an R terminal and then accessed through a web browser, hosted as a web service, or launched through an R IDE (for example, R Studio). The application can be run with or without an internet connection apart from loading online base maps.

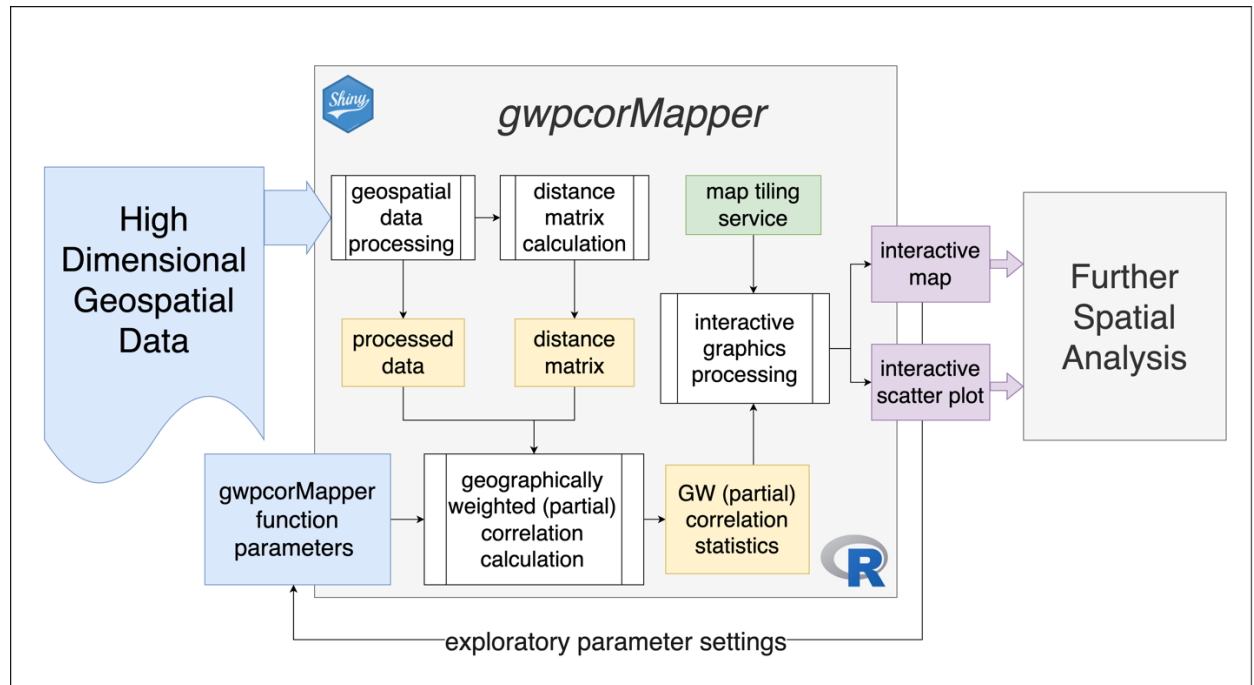

**Fig. 1**. The *gwpcorMapper* application diagram. Input features are colored blue, online features are colored green, key objects that are stored in memory are colored yellow, and output features are colored purple. The application flow is displayed through an iterative process that flows from left to right, where high-dimensional input data are added to the application with function parameters, the data is processed by *gwpcorMapper*, which then outputs interactive plots and data that can be explored and then used for successive spatial analysis.

We build *gwpcorMapper* from the extensive geospatial statistical libraries that R offers as open source on popular repositories like GitHub and CRAN. In particular, the application makes use of the powerful graphics and rendering capabilities of *plotly* (Sievert 2020), spatial data manipulation with *sf* (Pebesma 2018), and high-performance computation with *Rcpp* (Eddelbuettel and François 2011). Data import and spatial manipulation functionality are leveraged from the *sf* package to allow comprehensive geospatial data support, enabling *gwpcorMapper* to read from a variety of data types and provide the necessary spatial transformations for map rendering in a web application. *sf* in combination with *plotly* allows for interactive graphics and efficient rendering of the results of GW correlation and GW partial correlation. Such allows users to highlight selected detailed statistical and geospatial information about each plotted data point. The core algorithm logic for calculating GW correlation and GW partial correlation is written in C++ and ported to R using *Rcpp* to greatly enhance the performance of the application and in turn keep *gwpcorMapper* both lightweight and powerful.

2.2 GW Correlation and Partial Correlation
The core algorithm behind *gwpcorMapper* is based on R library functions from *GWpcor* (Percival and Tsutsumida 2017). *GWpcor*, which is an extension of the GW summary statistics function, *gwss()*, in the R library *GWmodel* (Lu et al. 2014; Gollini et al. 2015), provides GW correlation and GW partial correlation analyses with a statistical *t-test* to test the hypothesis that the correlation is zero. Like GW correlation in the *GWmodel* library, GW partial correlation calculates the weighted partial correlation between two variables while holding one or more variables constant under a moving kernel. Both Pearson's and Spearman's correlation types for either GW correlation or GW partial correlation are available in *gwpcorMapper*.

The calculations for both GW correlation and GW partial correlation begin by calculating the geographically weighted co-variance matrix ($\boldsymbol{\Sigma}$) at each data point *i*, whose location can be described by coordinates ($u_i, v_i$). Given a numeric matrix $\boldsymbol{P}$ of $n$ rows by $m$ variables such that $m_a$ and $m_b$ represents the a$^{th}$ and b$^{th}$ columns of $\boldsymbol{P}$ and each row represent the observations of each variable, we calculate the GW co-variance matrix as it is

$$\boldsymbol{\Sigma}(u_i, v_i) = \boldsymbol{P}^T \boldsymbol{W}(u_i, v_i) \boldsymbol{P} \qquad (1)$$

described by Harris, Brunsdon, and Charlton (2011):
where $\boldsymbol{P}$ is the data matrix containing the variables of interest and $\boldsymbol{W}(u_i, v_i)$ is the diagonal matrix of geographic weights across each location ($u_i, v_i$) that is created using one of the five kernel functions displayed in Fig. 2. Following this, either the GW correlation coefficients or the GW partial correlation coefficients are calculated. Defining $\boldsymbol{M}$ as the set of $m$ variables in $\boldsymbol{P}$, the GW correlation coefficient at any location ($u_i, v_i$) between two variables ($m_a$ and $m_b$) of $\boldsymbol{M}$ as described in Gollini et al. (2015) is given by:

$$\rho_{m_a m_b}(u_i, v_i) = \frac{\boldsymbol{\Sigma}_{m_a m_b}(u_i, v_i)}{s_{m_a}(u_i, v_i) s_{m_b}(u_i, v_i)} \qquad (2)$$

where $\boldsymbol{\Sigma}_{m_a m_b}(u_i, v_i)$ is the element of GW co-variance matrix between variables $m_a$ and $m_b$ at the location ($u_i, v_i$) and both $s_{m_a}$ and $s_{m_b}$ are the GW standard deviations of variable $m_a$ and $m_b$, respectively whose calculations follow Harris and Brunsdon (2010). Then, following Percival and Tsutsumida (2017), the GW partial correlation coefficients at any location ($u_i, v_i$) between two variables ($m_a$ and $m_b$) of the set $\boldsymbol{M}$, given all others in the set is given by:

$$\rho_{m_a m_b \cdot \boldsymbol{M} \backslash \{m_a, m_b\}}(u_i, v_i) = \frac{\boldsymbol{C}_{m_a m_b}}{\sqrt{\boldsymbol{C}_{m_a m_a} \boldsymbol{C}_{m_b m_b}}} \qquad (3)$$

where $\boldsymbol{C}$ is the inverted positive definite GW co-variance matrix between two variables ($m_a$ and $m_b$): $\boldsymbol{C} = (\boldsymbol{\Sigma}(u_i, v_i))^{-1}$, and $\boldsymbol{C}_{m_a m_b}$ is the element between variables $m_a$ and $m_b$. In the case that $\boldsymbol{\Sigma}(u_i, v_i)$ is not positive definite and therefore not invertible, the pseudo-inverse is calculated and applied using the Moore-Penrose inverse (Penrose 1955; Schafer et al. 2017). Finally, to find either Spearman's GW correlation coefficient or GW partial correlation coefficient, the variables of interest are ranked prior to the calculation of the GW covariance matrix (see Gollini et al. (2015) for details). The GW partial correlation and GW correlation coefficients can then be mapped to observe spatial heterogeneity in the association between two or more variables in spatial data.

**Table 2.** Kernel functions used in *gwpcorMapper* to calculate the diagonal matrix of geographic weights, $\boldsymbol{W}$. $\omega_{ij}$ is geographic weight at the $j^{th}$ element in $\boldsymbol{W}$ that is calculated for the $i^{th}$ observation in the data matrix, $\boldsymbol{P}$. $d_{ij}$ is the Euclidean distance between the $j^{th}$ observation point from $i$ and $b$ is the effective bandwidth.

| Name | Equation |
|---|---|
| Gaussian | $\omega_{ij} = \exp\left(-\frac{1}{2}\left(\frac{d_{ij}}{b}\right)^2\right)$ |
| Exponential | $\omega_{ij} = \exp\left(-\frac{|d_{ij}|}{b}\right)$ |
| Box-Car | $\omega_{ij} = \begin{cases} 1 & \text{if } |d_{ij}| < b \\ 0 & \text{otherwise} \end{cases}$ |
| Bi-Square | $\omega_{ij} = \begin{cases} \left(1-\left(\frac{d_{ij}}{b}\right)^2\right)^2 & \text{if } |d_{ij}| < b \\ 0 & \text{otherwise} \end{cases}$ |
| Tri-Cube | $\omega_{ij} = \begin{cases} \left(1-\left(\frac{|d_{ij}|}{b}\right)^3\right)^3 & \text{if } |d_{ij}| < b \\ 0 & \text{otherwise} \end{cases}$ |

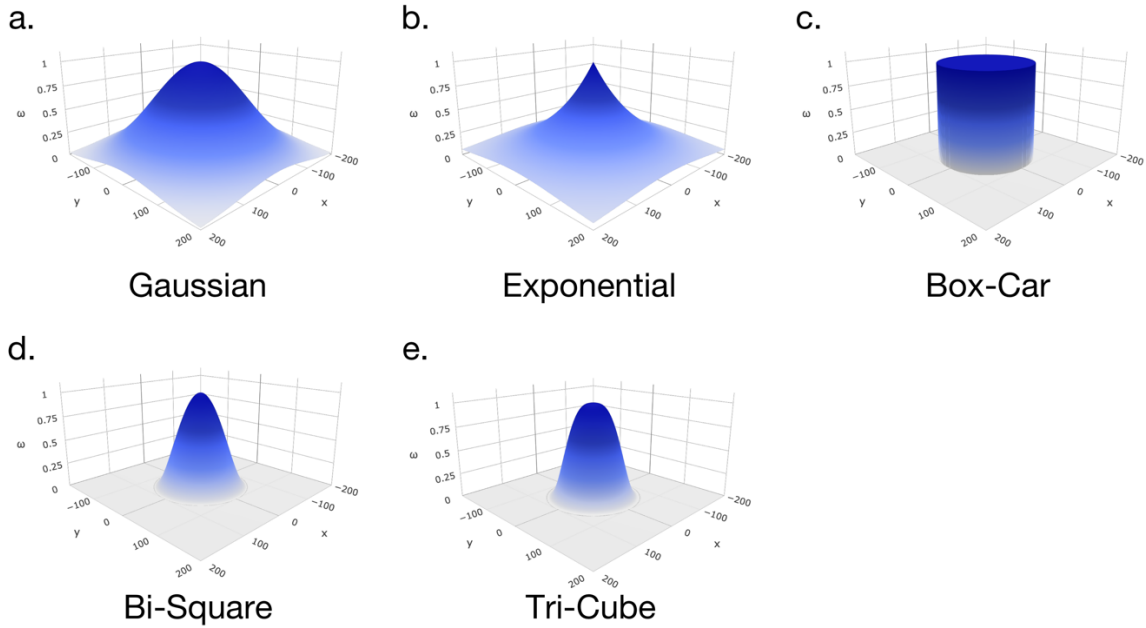

**Fig. 2.** Kernel functions available in *gwpcorMapper*: (a) Gaussian, (b) Exponential, (c) Box-Car, (d) Bi-Square, and (e) Tri-Cube. Spatial weights are defined by each respective kernel whose functions are given in Table 2.

### 2.3 Software Usage
*gwpcorMapper* can read several popular geospatial file types, including geopackage, geojson, and ESRI shapefiles, as input and accepts several input fields as parameters (Table 2, Fig. 1 and 3). Users of the software can also specify base map layers provided by map tiling services such as CARTO (CARTO 2021) and Mapbox (Mapbox 2021) or by providing the layer source URL of a self-hosted solution like GeoServer (Open Source Geospatial Foundation, Geoserver Contributors 2021). The *gwpcorMapper* interface consists of three panels: (I) the data panel, (II) the map panel, and (III) the scatter plot panel (Fig. 3). The data panel contains the controls for loading data and setting parameters (input), while the map and scatter plot panels display the resultant correlation coefficients on both a map and a scatter plot respectively (output). The map is displayed as either a choropleth or point map depending on the geometry type of the input data. The map and the plot are themselves interactive by providing the user with linked views between the two. For example, selection of any plot element in the map will highlight the same element in the scatter plot (and vice versa) and will provide the correlative statistics at that point. Both the map and scatter plots use a continuous color palette based on the *viridis* color scheme (Garnier 2018).

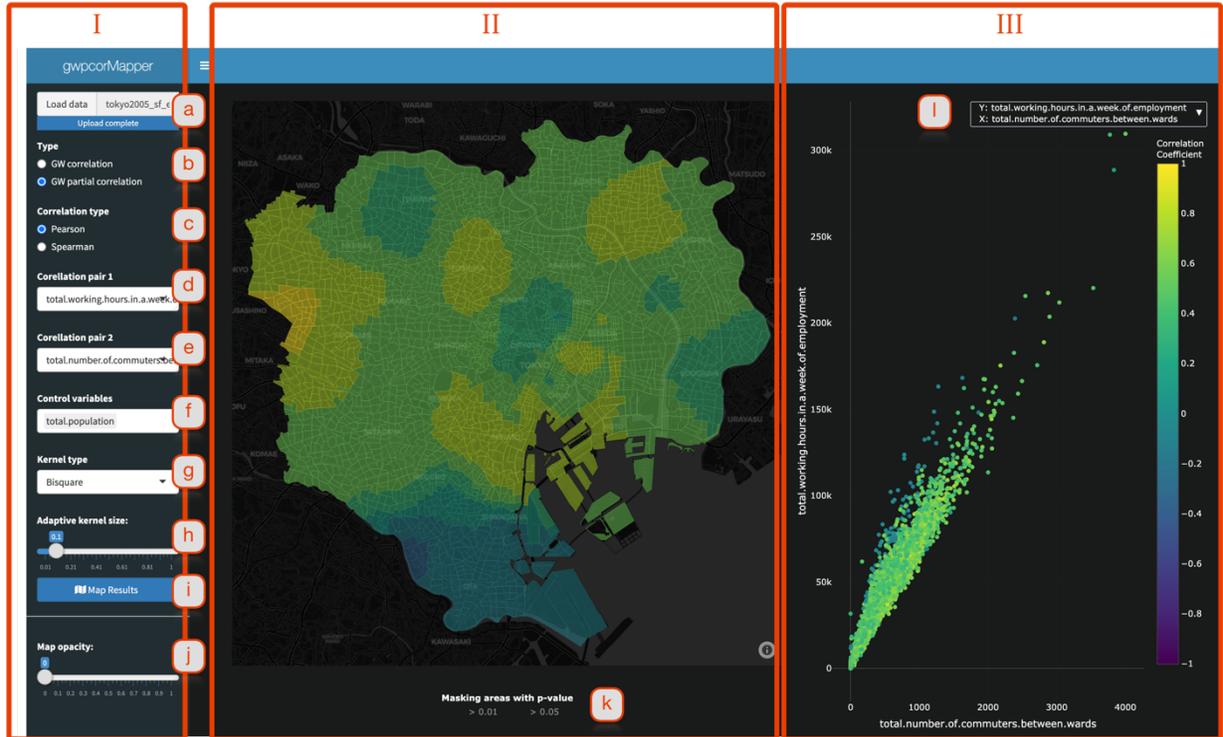

**Fig. 3.** The *gwpcorMapper* UI: a multi-panel user interface highlighting its core features: (I) parameter selection, (II) map, and (III) scatter plots. Input features are labelled with letters (a-l) for cross reference with Table 3.

**Table 3.** *gwpcorMapper* input fields, their descriptions, and cross-reference label to Fig. 3

| Input Field | Description | Label |
|---|---|---|
| Load Data | Upload component that allows users to select data files for analysis. | a |
| Type | Radio selectors for GW model. Users can choose between GW correlation and GW partial Correlation. | b |
| Correlation Type | Radio selectors for the correlation method to be employed. Users can choose between Pearson and Spearman. | c |
| Correlation Pair 1 | Searchable drop-down field to select first variable for analysis. | d |
| Correlation Pair 2 | Searchable drop-down field to select second variable for analysis. | e |
| Control Variables | Searchable drop-down field to select any number of control variables for analysis. | f |
| Kernel Type | Drop-down field to select kernel type for analysis. Available kernel types are described in Fig. 2. | g |
| Adaptive Kernel Size | Bandwidth selector for kernel size in GW analysis. Only adaptive kernels are currently supported. | h |
| Map Results | Button to trigger analysis and to map the results. | i |
| Map Opacity | Opacity control of results layer on map. | j |
| *P* value Mask | Buttons to trigger masking statistical *t-test* using *P* values at the 0.01 or 0.05 thresholds. | k |
| Variable Selector | Drop-down field to select which variable pair to plot in partial correlation analysis. | l |

## 3. Case Study

### 3.1 Data

Data from the national census for 2005 and the economic census for 2006 of the 23 special wards in Tokyo, Japan, at *chocho-aza* level (a total of 3134 observations) is prepared to demonstrate an example of ESDA. The administrative boundaries of the wards are observed by using *gwpcorMapper* to load a custom base map. The census data consists of 228 variables consisting of basic demographic data including total population broken down by age, gender, employment status, and other basic socio-economic indicators. We use *gwpcorMapper* to study the urban social structure of the 23 special wards in Tokyo by selecting various covariates and mapping both GW correlation and GW partial correlation with various parameter settings. This allows us to explore the data to find meaningful relationships that describe the work-life environment in Tokyo, and then discuss our findings with respect to the compact city concept.

### 3.2 ESDA Procedure with gwpcorMapper

Our first step in ESDA with *gwpcorMapper* is variable selection. As our interest is to know the work-life environment, we selected the following variables to describe the exploratory process with *gwpcorMapper* in this paper: 1. The "total number of commuters between wards" (herein referred to as "*commuters*"), 2. The "total number of households" (herein referred to as "*households*"), 3. The "*day-night population ratio*", 4. The "average area per person" (herein referred to as "*area per person*"), 5. The "total working hours in a week of employment" ("*working hours*"), 6. The "*day time population*", and 7. The "*total population*". Many other variables can be selected but we limit 7 variables only in this paper to demonstrate the ESDA procedure with *gwpcorMapper* simply.

After variable selection, we then observe any spatial heterogeneity in the correlation between these variables by first observing the GW correlation between bivariate pairs and then GW partial correlation. Following this, we demonstrate how we can examine statistical significance of these relationships by examining *p*-values on resulting maps. Finally, we pick out the relationship between *commuters* and *working hours* while controlling for *total population* to observe how the spatial patterns of the relationship copes under varying bandwidths and kernel functions. Such may inform successive studies by urban planning experts on how the diversified urban social structure of compact cities may influence working hours and the need to commute further distances to work.

### 3.3 Case Study Results

Examination of the seven selected census data variables reveals that there are a total of 21 bivariate pairs alone for analysis at each of the 3134 administrative regions. The ability to search and select variables from drop-down menus and have the results of GW correlation and partial correlation mapped on the fly enabled us to quickly see how certain variables may have more pronounced spatial variation than others.

#### 3.3.1 GW Correlation Coefficients

Fig. 4 displays 6 of the 21 resulting maps of the GW correlation with *gwpcorMapper* using an adaptive bandwidth of 0.25 (25% of the total spatial units falls into a kernel) and a Bi-Square kernel function. These demonstrate a range of spatial patterns in the relationship between each pair; from the seemingly strong spatially homogenous correlation between both *commuters* and *households* (Fig. 4.a) and *commuters* and *working hours* (Fig. 4.f), to the weak negative spatially heterogenous correlation between both *commuters* and *day-night population ratio* (Fig. 4.b) and *working hours* and *day-night population ratio* (Fig. 4.d), and to the spatially heterogenous correlation between *commuters* and *area per person* (Fig. 4.c) and *day-time population* with *commuters* (Fig. 4.e).

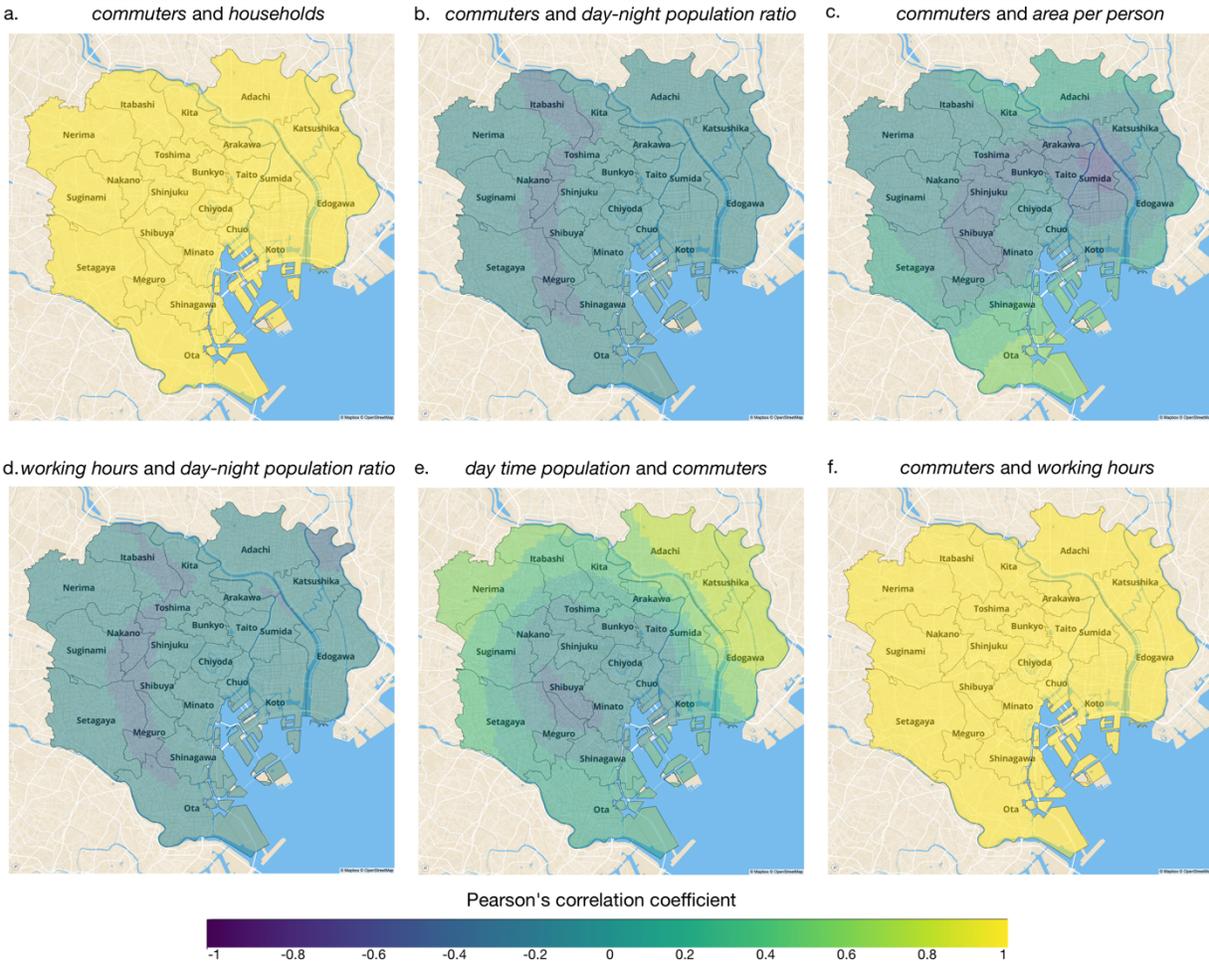

**Fig 4.** Map panel output showing GW correlation between: a. the total number of commuters between wards and the total number of households, b. the total number of commuters between wards and the day-night population ratio, c. the total number of commuters between wards and the average area per person , d. the total working hours in a week of employment and the day-night population ratio, e. the day time population and the total number of commuters between wards , and f. the total number of commuters between wards and the total working hours in a week of employment.

3.3.2 GW Partial Correlation Coefficients
To demonstrate the functionality of *gwpcorMapper* to explore the effects that changes in additional variables may have on the correlation between any two variables, we map GW partial correlation between each of the variable pairs mapped in Fig. 4 while controlling for *total population* without changing any bandwidth parameters. Fig. 5 displays how using GW partial correlation to control for a third (or more) variable can drastically change the spatial patterns between each relationship. Notably, we can see that strong patterns emerge between both *commuters* and *households* (Fig. 5.a), and *commuters* and *working hours* (Fig. 5.f) while controlling for *total population*. In Fig. 5.a we can see that there is a tendency for districts in wards that are highly residential, like Nerima, Itabashi, and Adachi, to display a strong positive correlation between *households* and *commuters*, while highly commercial areas like Minato, Chuo, and Chiyoda display a strong negative correlation. Interestingly, we see less of a pronounced relationship between *commuters* and *day-night population ratio* (Fig. 5.b), *commuters* and *area per person* (Fig. 5.c), *working hours* and *day-night population ratio* (Fig. 5.d), and *day-time population* and *commuters* (Fig. 5.e) once we control for the effects that changes in total population may have on each relationship across the 23 special wards.

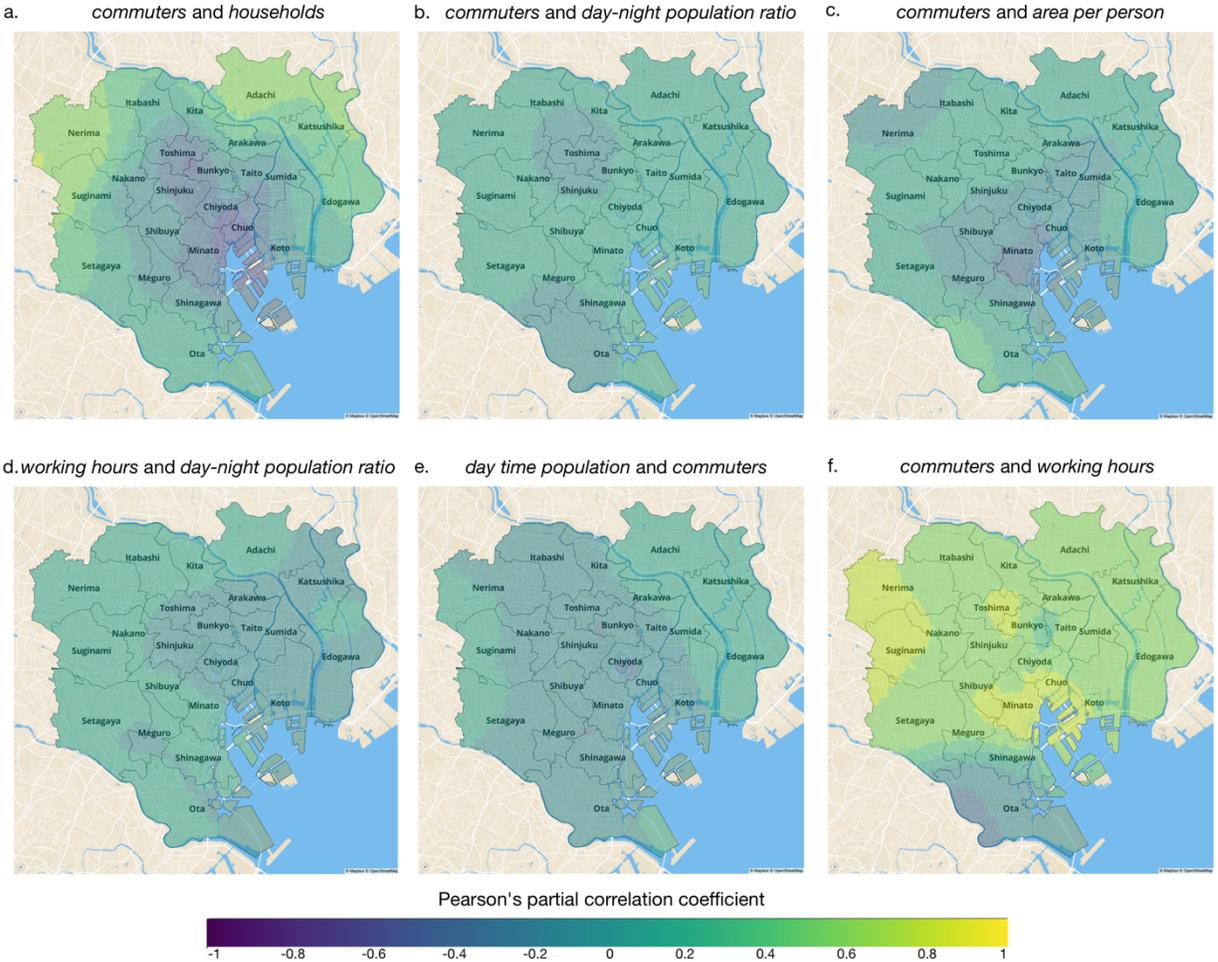

**Fig. 5.** Map panel output showing GW partial correlation between: a. the total number of commuters between wards and the total number of households, b. the total number of commuters between wards and the day-night population ratio, c. the total number of commuters between wards and the average area per person, d. the total working hours in a week of employment and the day-night population ratio, e. day time population and the total number of commuters between wards, f. the total number of commuters between wards and the total working hours in a week of employment, while controlling for total population.

### 3.3.3 Statistical Significance Masking

The weakening and levelling out of the correlations displayed in Fig. 5.b through 5.e that is apparent once we control for *total population* suggest that these pairs may not lead to any further interesting analyses. Displaying these correlations while masking statistically insignificant districts (with statistical significance being defined where *p*-value $\leq 0.01$) provide further evidence to suggest these decisions. In Fig. 6, we can see that the relationships between *commuters* and *households* (Fig. 6.a) and *commuters* and *working hours* (Fig. 6.f), while controlling for *total population*, display reasonably clear relationships; yet the relationship between *commuters* and *day-night population ratio* (Fig. 5.b), *and working hours* and *day-night population ratio* (Fig. 5.d), do not.

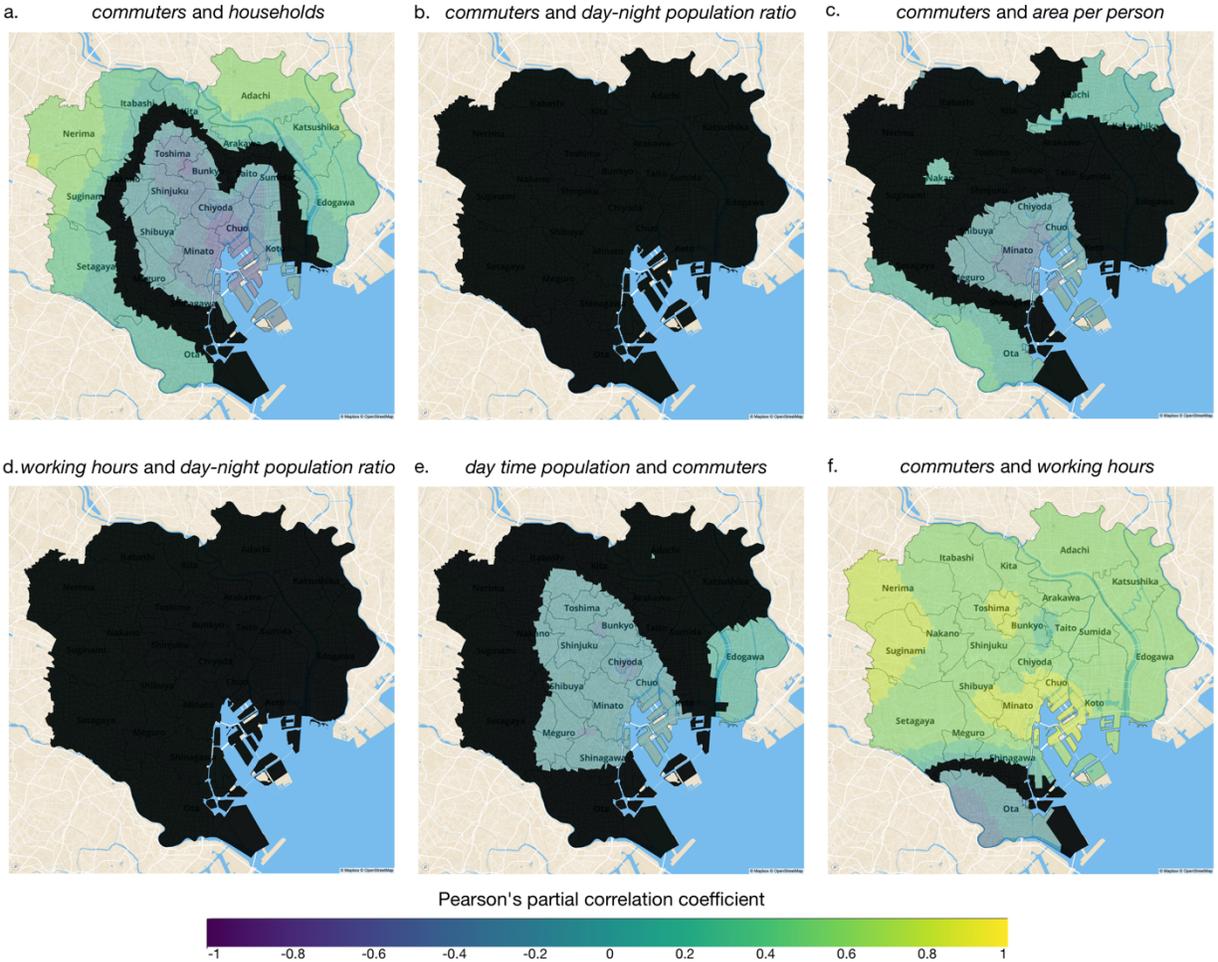

**Fig. 6.** Map panel output showing statistical insignificance masking using *p* values for the GW partial correlation between: a. the total number of commuters between wards and the total number of households, b. the total number of commuters between wards and the day-night population ratio, c. the total number of commuters between wards and the average area per person, d. the total working hours in a week of employment and the day-night population ratio, e. day time population and the total number of commuters between wards, f. the total number of commuters between wards and the total working hours in a week of employment, while controlling for total population. Statistical significance is defined where *p* values are ≤ 0.01.

### 3.3.4 Bandwidth and Kernel Functions

Fig. 4, 5, and 6 all displayed either GW correlation or GW partial correlation using Pearson's correlation statistic with a constant adaptive bandwidth of 0.25 and a Bi-Square kernel function. However, further examination of the maps under varying bandwidth sizes and kernel types reveals differences in the degree of spatial heterogeneity of the (partial) correlation coefficient and their resulting spatial patterns. Focusing on the relationship between *commuters* and *working hours* while controlling for *total population* and keeping the application of a Bi-Square kernel enables us to see how changing the size of the bandwidth can exaggerate the degree of spatial heterogeneity. Very large bandwidths lead to significantly reduced spatial variation with correlation coefficients levelling to moderate and positive values. Smaller bandwidths reveal that there are areas where the correlation statistics are widely different across localities.

Using an adaptive bandwidth of 0.1, we can see in Fig. 7.a that there are some areas that display a strongly positive relationship between *commuters* and *working hours*, such as the highly residential areas surrounding Suginami and Toshima wards or the central downtown area of Minato; some, like the central commercial ward of Chiyoda, that only display a very weakly positive one; and others, like the Ota ward where the presence of large factories with on-board housing drives lower commuting needs, display a negative correlation between *commuters* and *working hours*. Larger bandwidths (an adaptive bandwidth of 0.5) even out the local events and most localities display a uniform weak-to-moderate positive correlation with more gradual spatial variation. We still see the strong positive relation in Suginami and Minato, however there are no longer any patches of strong correlation in Chiyoda, Shibuya, or Sumida. We also no longer see the weaker correlations in Edogawa and Itabashi. There are consistent patterns within certain areas like Koto but not in others, like Chuo. Thus, using *gwpcorMapper* to explore the data, we can explore the change and the consistency of the correlation between *commuters* and *working hours* according to the degree of spatial heterogeneity. It is difficult to achieve such findings using existing available applications.

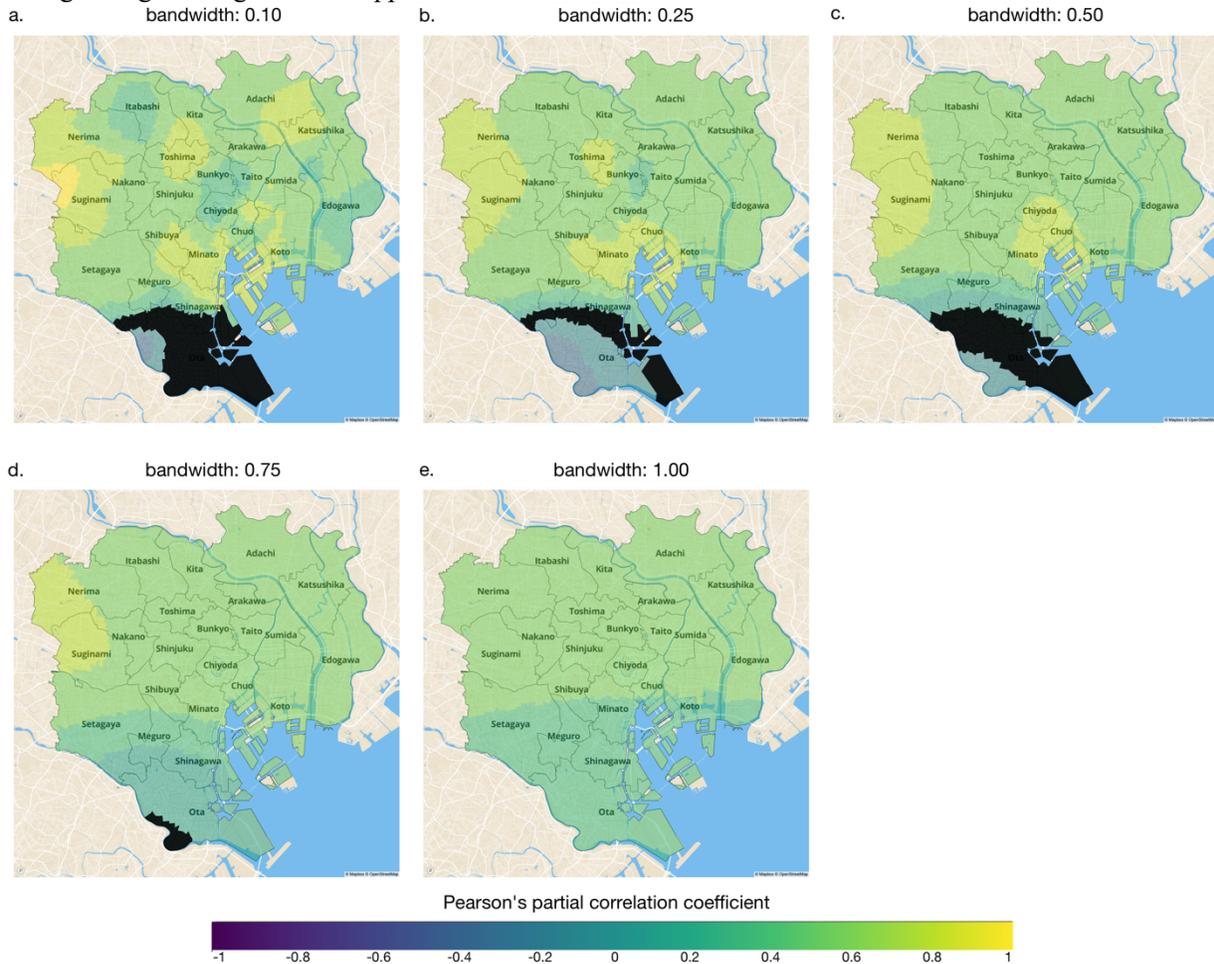

**Fig. 7.** Resulting maps of GW partial correlation coefficients between commuters and working hours while controlling for total population using a Bi-Square kernel and apply varying adaptive bandwidth sizes (in terms of proportions of the total data): (a) 0.1, (b) 0.25, (c) 0.5, (d) 0.75, (e) 1.0. Non statistically significant results at the 0.01 level are masked in black.

Finally, we can observe the effects of the weighting scheme in the GW correlation and partial correlation by using different kernel functions using the adaptive bandwidth of 0.1 (Fig. 8). Up until this point, we have been using a Bi-Square kernel to apply spatial weights to the data. However, there are four other kernel functions available in *gwpcorMapper*. Resultant maps by Gaussian and Exponential kernel functions which apply weights to all observations display gradual spatial variations of the correlative relationship while maps made using Box-Car, Bi-Square, and Tri-Cube kernel functions, which apply weights to only observations that fall within the bandwidth size, display distinct localized variations of the relationship. These local patterns describe interesting features of the data and suggest that any of the piece-wise kernel functions may be more appropriate for this data.

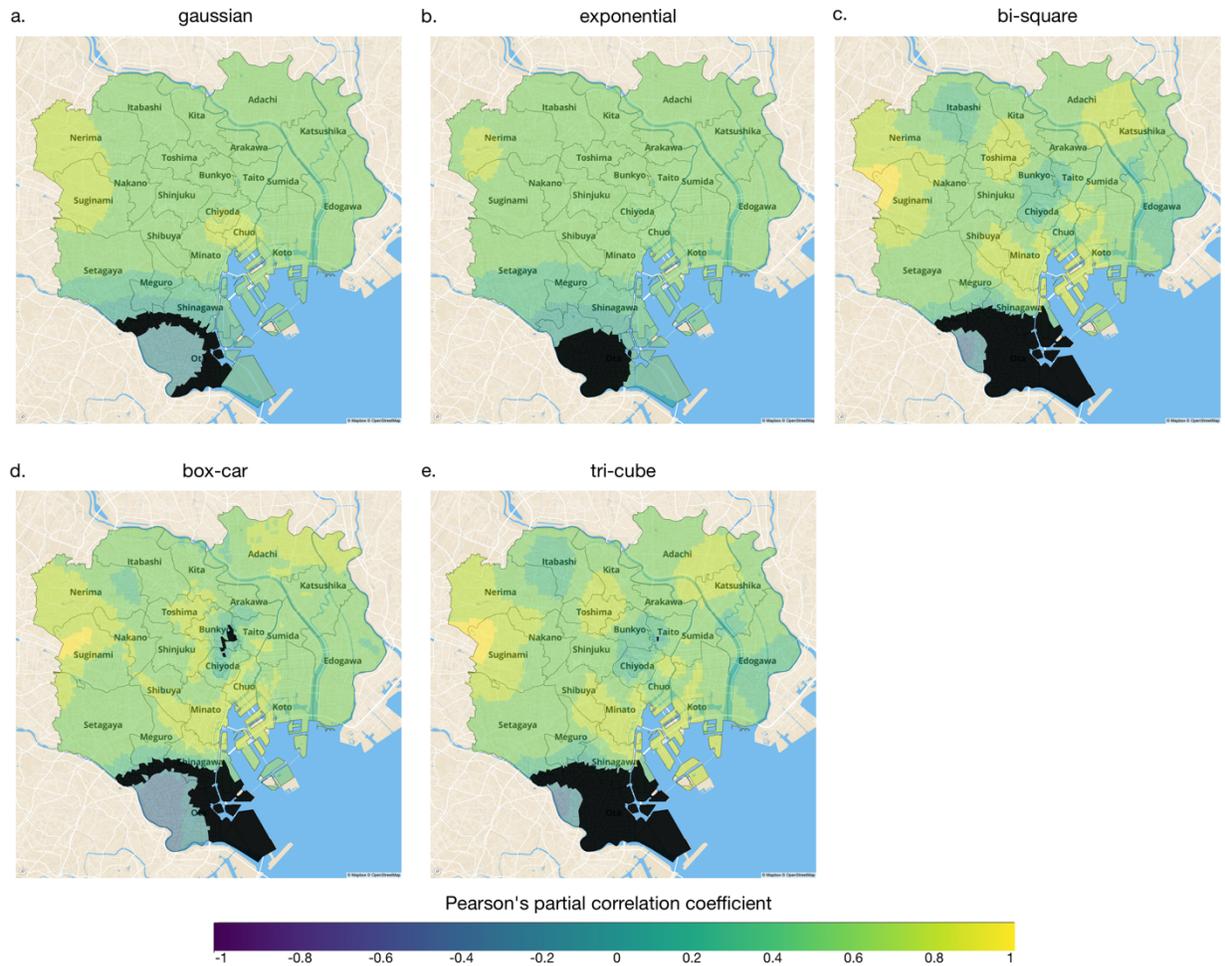

**Fig. 8.** Resulting maps of GW partial correlation coefficients between commuters and working hours while controlling for total population using a constant (proportional) bandwidth size of 0.1 while applying varying kernel types: (a) Gaussian, (b) Exponential, (c) Bi-Square, (d) Box-Car, and (e) Tri-Cube. Non statistically significant results at the 0.01 level are masked in black.

3.4 Computational Performance

As a software that conducts ESDA, it is important that the analysis of *gwpcorMapper* be as fast and as efficient as possible. Especially when there can be a very high number of combinations between variables and model parameters. We benchmarked the performance of *gwpcorMapper* to calculate GW correlation statistics (without data loading and map rendering) and compared these with three other software libraries that calculate GW correlation statistics: *GWmodel* (v2.2-8), *spgwr* (v0.6-34), and *lctools* (v0.2-8). In each test we used an adaptive bandwidth of 0.2, a bi-square kernel function, and Pearson's correlation statistic. We made benchmark tests using both the case study dataset, selecting "*total population*", "*day time population*", and "*population density*" as variables for analysis (Table 4), as well as real estate housing data from Zillow to benchmark the scalability of *gwpcorMapper* under datasets of increasing size (Fig. 9). The Zillow dataset comprises of random samples of housing property data in Los Angeles that was first published on Kaggle.com as a part of a competition to improve Zillow's property value estimation algorithm. This data has also been used in other research (for example, Li et al. 2019) to benchmark performance in GW models and the data we use consists of sub-samples taken from the dataset published on GitHub by Li et al. (2019). The case study dataset consisted of 3134 observations while the Zillow datasets ranged between 100 to 50,000 observations. All analyses were performed on a MacBook Pro equipped with a 2.3 GHz 8-Core Intel Core i9 processor and 16GB of RAM. As demonstrated in Table 4 and Fig. 9, *gwpcorMapper* was orders of magnitude faster and less memory intensive than other similar R libraries across a high number of observations.

Table 4. Computational benchmarks and performance comparisons between *gwpcorMapper* and three other popular R package functions that can calculate GW Correlation statistics using the case study dataset which consists of 3024 observations.

|                  | CPU Time (s) | Memory Allocated (MB) |
|------------------|--------------|------------------------|
| GWmodel::gwss    | 810.470      | 16200                  |
| lctools::lcorrel | 73.069       | 21700                  |
| spgwr::gw.cov    | 10.753       | 5360                   |
| gwpcorMapper     | 1.776        | 584                    |

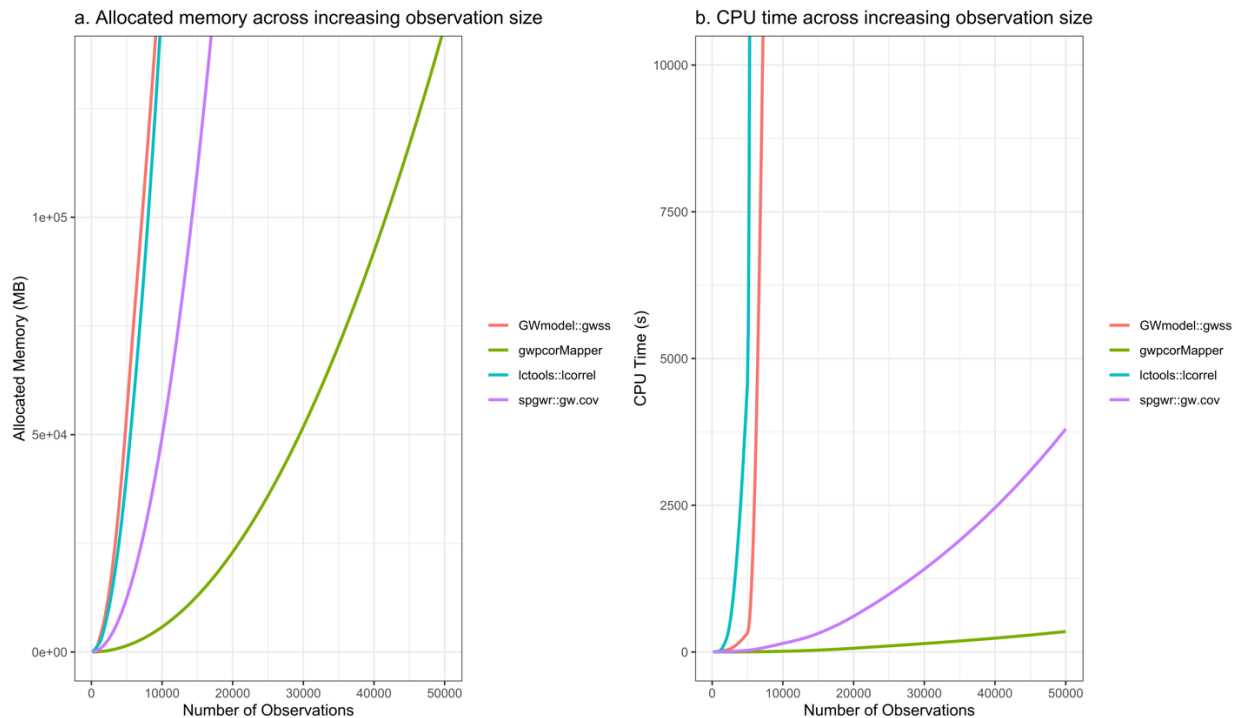

Figure 9. Performance benchmarks of *gwpcorMapper* and existing software across datasets of increasing size in terms of a. allocated memory (in megabytes) and b. CPU time (in seconds). This was tested on a MacBook Pro equipped with a 2.3 GHz 8-Core Intel Core i9 processor and 16GB of RAM.

## 4. Discussion

*gwpcorMapper* enabled us to work through a series of exploratory steps to investigate spatial heterogeneity interactively using geographically weighted correlation and partial correlation analyses across multiple variables in a high-dimensional dataset. In our case study, we found that observing the differences in the degree of association between *commuters* and *working hours* once we control for *total population* provided a strikingly simple and important example of ESDA using *gwpcorMapper*.

It is obvious that there is a very homogenous, high, and tight association between the total number of commuters in a ward and that ward's total working hours of its citizens. Both are likely driven by the total number of residents in each ward, where wards with a higher total number of people are more likely to have more commuters and a larger total number of hours worked in employment. Indeed, mapping the correlation coefficients of *total population* with either *commuters* or *working hours* will reveal an equally spatially stationary strong positive relationship. Thus, adding the total population of a ward as a controlling variable provided a much more realistic representation of the degree of association between these two variables. Such spatial association was revealed and varied according to the bandwidth parameters (size and shape), and we could explore spatial changes and consistencies by the degree of spatial heterogeneity of the association through *gwpcorMapper*. Such findings emphasize the importance of the explorative investigation of data prior to developing further analyses such as modeling, and policy making.

The graphical user interface and resulting maps from *gwpcorMapper* allowed for the exploration of variable and parameter combinations in a simple and fast manner. Using *gwpcorMapper* we could select a subset of variables and explore a range of parameters, tuning as we went a long, to find meaningful relationships between our covariates. The interactive map, scatter plots, statistical significance masking allow analysts to zoom into features of interest and highlight any noteworthy observations. Working in *gwpcorMapper* provides the advantage of being able to see results immediately after making variable choices. It gives analysts the freedom and control to tune and adjust parameter decisions based on immediate results. As it is demonstrated in Fig. 9, not only would it take up to 100-times more time to simply make the GW correlation calculation using existing software like *GWmodel*, but analysts would also need to write code for the program to make these plots. Any adjustments in parameters would require minor modifications to the code base and parameter search would become an exhaustive operation with potentially thousands of figures plotted for comparison.

4.1 Interpreting the Case Study
Under the lens of the compact city framework, our results from ESDA suggest that either having to commute further distances may encourage longer working hours, or simply that jobs with longer working hours are concentrated in the city center where many workers need to commute to. If working hours can be reduced by removing the need to commute long distances, then workers can spend more time doing other things. If commuting is correlated with longer working hours, then on top of losing time during commute, people have even less time to perform social functions due to their longer working hours. These concepts will be investigated by further analyses and support recent work where GWR was used to suggests that worker sleep problems within cities may be related to longer commuting times (Kim et al. 2019).

4.2 Future Improvements
It is important to note that our case study involved the mapping of bivariate pairs between 7 variables while controlling for 1 variable in each partial correlation map. Such variable selection need not be limited to such a small selection. An interesting area for further development may be for the software to suggest "potential worthwhile relationships for further investigation" in the entire high-dimensional data. Such an extension may be used to dramatically reduce the number of pairs that are necessary for analysis and can be addressed in future development work. Another potential improvement is to incorporate additional interactive features, such as a map compare function that will allow users to compare the results of two or more analyses through interactive side-by-side maps.

5. Conclusions
*gwcorMapper* is applicable to many popular geospatial data formats, and it is not limited to census data analysis in the urban planning domain as shown in our case study. GW frameworks have been used to explore and analyze a wide variety of phenomena, from applications in spatial econometrics and epidemiology, to ecology and climatology. For example, GWR has been applied to describe housing market dynamics (Yu, Wei, and Wu 2007), to provide better estimations of fish biomass (Irigoien et al. 2014), to predict allometric relations in trees (Guo, Ma, and Zhang 2008), to understand the effects of local conditions on atmospheric $PM_{2.5}$ concentrations (Luo et al. 2017), and to predict the occurrence of hand foot and mouth disease (Hong et al. 2018). *gwpcorMapper* provides the tools for exploring spatial heterogeneity in the data prior to analysis and can be used for initial ESDA in any of these scenarios.

*gwpcorMapper* is open source and its source code along with installation and usage instructions can be found on GitHub at https://github.com/gwpcor/gwpcormapper. It is also available as the R package "gwpcormapper" on CRAN and can be installed in R using the command: `install.packages("gwpcormapper")`. The features and results of *gwpcorMapper* presented in this paper were made with version 0.1.3.

**Competing Interests**
All authors do not have any competing interests which should be declared for this submission.

**Funding**
This research was funded by the Joint Support Center for Data Science Research at Research Organization of Information and Systems (ROIS-DS-JOINT) under Grant 006RP2018, 004RP2019, 003RP2020, and 005RP2021.

**Author Contributions**
NT conceived the original idea behind *gwpcorMapper* and both JEHP and NT developed the software with JEHP being the main contributor. JEHP prepared the initial draft of the manuscript and produced the visualizations. All authors contributed to the writing of the paper. All authors have read and agreed to the published version of the manuscript.

**Data Availability Statement**
https://cran.r-project.org/web/packages/gwpcormapper/index.html

**Conflicts of Interest**
The authors declare no conflict of interest. The funders had no role in the design of the study; in the collection, analyses, or interpretation of data; in the writing of the manuscript, or in the decision to publish the results.

**References**
Anselin, L (1988a) Spatial Econometrics: Methods and Models. Stud Oper R. https://doi.org/10.1007/978-94-015-7799-1.
Anselin L (1988b) Lagrange Multiplier Test Diagnostics for Spatial Dependence and Spatial Heterogeneity. Geogr Anal. 20:1-17 https://doi.org/10.1111/j.1538-4632.1988.tb00159.x
Anselin L, Rey S (1991) Properties of Tests for Spatial Dependence in Linear Regression Models. Geogr Anal. 23:112-131 https://doi.org/10.1111/j.1538-4632.1991.tb00228.x
Anselin L (1999) The Future of Spatial Analysis in the Social Sciences. Ann Gis. 5:67-76 https://doi.org/10.1080/10824009909480516
Anselin L (2005) Interactive Techniques and Exploratory Spatial Data Analysis. In P. A. Longley, M. F. Goodchild, D. J. Maguire, and D. Rhind (eds) Geographical Information Systems: Principles, Techniques, Management and Applications, 2nd edn. Wiley, pp 253–266
Anselin L, Syabri I, Kho Y (2006) GeoDa: An Introduction to Spatial Data Analysis. Geogr Anal. 38:5-22 https://doi.org/10.1111/j.0016-7363.2005.00671.x
Bibri S E, Krogstie J Kärrholm M (2020) Compact City Planning and Development: Emerging Practices and Strategies for Achieving the Goals of Sustainable Development. Dev Built Environ. https://doi.org/10.1016/j.dibe.2020.100021
BioMedware (2020) Spacestat. https://www.biomedware.com/software/spacestat/. Accessed 22 February 2021
Bivand R S (2010) Exploratory Spatial Data Analysis. In Fishcer M and Getis A (eds) Handbook of Applied Spatial Analysis: Software Tools, Methods and Applications. Springer, Heidelberg, pp 219–254
Bivand R, Yu D (2020) spgwr: Geographically Weighted Regression. R package version 0.6-34. https://CRAN.R-project.org/package=spgwr. Accessed 22 February 2021
Brunsdon C, Fotheringham A S, Charlton M E (1996) Geographically Weighted Regression: A Method for Exploring Spatial Nonstationarity. Geogr Anal 28(4):281–298. https://doi.org/10.1111/j.1538-4632.1996.tb00936.x
Brunsdon C (1998) Exploratory spatial data analysis and local indicators of spatial association with XLISP-STAT. J Royal Statistical Soc Ser D Statistician. 47(3):471–484. https://doi.org/10.1111/1467-9884.00148
Brunsdon C, Fotheringham A S, Charlton M (2002) Geographically weighted summary statistics — a framework for localised exploratory data analysis. Comput Environ Urban Syst. 26(6):501–524. https://doi.org/10.1016/s0198-9715(01)00009-6

Brunsdon C, Fotheringham S, Charlton M (2007) Geographically Weighted Discriminant Analysis. Geogr Anal. 39(4):376–396. /https://doi.org10.1111/j.1538-4632.2007.00709.x

CARTO (2021) CARTO. https://carto.com/. Accessed 25 February 2021

Chang W, Cheng J, Allaire JJ, Xie Y, McPherson J (2020) Shiny: Web Application Framework for R. R package version 1.5.0. https://CRAN.R-project.org/package=shiny. Accessed 4 December 2020

Cliff A D, Ord J K (1973) Spatial Autocorrelation. Pion, London.

Comber A, Brunsdon C, Charlton M, Dong G, Harris R, Lu B, LuY, Murakami,D, Nakaya T, Wang Y, Harris P (2022) A Route Map for Successful Applications of Geographically Weighted Regression. Geogr Anal. https://doi.org/10.1111/gean.12316

Dantzig G, Saaty T L (1973) Compact City: a Plan for a Livable Urban Environment. W. H. Freeman.

Dykes J (1998) Cartographic Visualization. J Royal Statistical Soc Ser D Statistician. 47(3):485–497. https://doi.org/10.1111/1467-9884.00149

Dykes J, Brunsdon C (2007) Geographically Weighted Visualization: Interactive Graphics for Scale-Varying Exploratory Analysis. IEEE Transactions on Visualization and Computer Graphics. 13(6):1161–1168. https://doi.org/10.1109/tvcg.2007.70558

Eddelbuettel D, François R (2011) Rcpp: Seamless R and C++ Integration. J Stat Softw. 40(8). https://doi.org/10.18637/jss.v040.i08

ESRI (2020) ArcGIS Desktop: Release 10.8. Redlands, CA: Environmental Systems Research Institute. https://www.esri.com/en-us/arcgis/about-arcgis/overview. Accessed 22 February 2021

Farber S, Páez A (2007) A systematic investigation of cross-validation in GWR model estimation: Empirical analysis and Monte Carlo simulations. J Geogr Syst. 9(4):371-396. https://doi.org/10.1007/s10109-007-0051-3

Fotheringham A (1992) Exploratory Spatial Data Analysis and GIS. Environment and Planning A, 12(24):1675–1678.

Fotheringham A S, Charlton M, Brunsdon C (1997) Recent Developments in Spatial Analysis, Spatial Statistics, Behavioural Modelling, and Computational Intelligence. Adv Spat Sci. 60–82. https://doi.org/10.1007/978-3-662-03499-6_4

Fotheringham S A, Brunsdon C, Charlton M (2003) Geographically Weighted Regression—The Analysis of Spatially Varying Relationships. John Wiley & Sons.

Garnier S (2018) viridis: Default Color Maps from 'matplotlib'. R package version 0.5.1. https://CRAN.R-project.org/package=viridis. Accessed 25 February 2021

Getis A (1994) Spatial dependence and heterogeneity and proximal databases. Spatial analysis and GIS. 105-120.

Gollini I, Lu B, Charlton M, Brunsdon C, Harris P (2015) GWmodel: An R Package for Exploring Spatial Heterogeneity Using Geographically Weighted Models. J Stat Softw. 63(17). https://doi.org/10.18637/jss.v063.i17

Guo L, Ma Z, Zhang L (2008) Comparison of bandwidth selection in application of geographically weighted regression: a case study. Can J Forest Res. 38(9):2526–2534. https://doi.org/10.1139/x08-091

Haining R, Wise S Ma J (1998) Exploratory Spatial Data Analysis. J Royal Statistical Soc Ser D Statistician. 47(3):457–469. https://doi.org/10.1111/1467-9884.00147

Harris P, Brunsdon C (2010) Exploring spatial variation and spatial relationships in a freshwater acidification critical load data set for Great Britain using geographically weighted summary statistics. Comput Geosci, 36(1): 54–70. https://doi.org/10.1016/j.cageo.2009.04.012

Harris P, Brunsdon C, Charlton M (2011) Geographically weighted principal components analysis. Int J Geogr Inf Sci. 25(10):1717–1736. https://doi.org/10.1080/13658816.2011.554838

Hong Z, Hao H, Li C, Du W, Wei L, Wang H (2018) Exploration of potential risks of Hand, Foot, and Mouth Disease in Inner Mongolia Autonomous Region, China Using Geographically Weighted Regression Model. Sci Rep-uk. 8(1):17707. https://doi.org/10.1038/s41598-018-35721-9

Irigoien X, Klevjer T A, Røstad A, Martinez U, Boyra G, Acuña J L, Bode A, Echevarria F, Gonzalez-Gordillo J I, Hernandez-Leon S, Agusti S, Aksnes D L, Duarte C M, Kaartvedt S (2014) Large mesopelagic fishes biomass and trophic efficiency in the open ocean. Nat Commun. 5(1):3271. https://doi.org/10.1038/ncomms4271

Kalogirou S (2020) lctools: Local Correlation, Spatial Inequalities, Geographically Weighted Regression and Other Tools. R package version 0.2-8. https://CRAN.R-project.org/package=lctools. Accessed 29 March 2022

Kim S, Kim Y, Lim S-S, Ryoo J-H, Yoon J-H (2019) Long Commute Time and Sleep Problems with Gender Difference in Work–Life Balance: A Cross-sectional Study of More than 25,000 Workers. Saf Heal Work, 10(4):470–475. https://doi.org/10.1016/j.shaw.2019.08.001

Lee S-I (2009) Neighborhood Effects. In Kitchin R, Thrift N (eds) International Encyclopedia of Human Geography. Oxford, Elsevier, pp 349–353

Li Z, Fotheringham S A, Li W, Oshan T (2019) Fast Geographically Weighted Regression (FastGWR): a scalable algorithm to investigate spatial process heterogeneity in millions of observations. Int J Geogr Inf Sci. 33(1):155-175. https://doi.org/10.1080/13658816.2018.1521523

Lloyd C D (2010) Analysing population characteristics using geographically weighted principal components analysis: A case study of Northern Ireland in 2001. Comput Environ Urban Syst. 34(5):389–399. https://doi.org/10.1016/j.compenvurbsys.2010.02.005

Lu B, Harris P, Charlton M, Brunsdon C (2014) The GWmodel R package: further topics for exploring spatial heterogeneity using geographically weighted models. Geo-spatial Information Sci. 17(2):85–101. https://doi.org/10.1080/10095020.2014.917453

Luo J, Du P, Samat A, Xia J, Che M, Xue Z (2017) Spatiotemporal Pattern of PM2.5 Concentrations in Mainland China and Analysis of Its Influencing Factors using Geographically Weighted Regression. Sci Rep-uk. 7(1):40607 https://doi.org/10.1038/srep40607

Mapbox (2021) Mapbox. https://www.mapbox.com/. Accessed 25 February 2021

Open Source Geospatial Foundation, GeoServer Contributors (2021) GeoServer; Version 2.18.2. http://geoserver.org/download/. Accessed on 25 February 2021

Pebesma E (2018) Simple Features for R: Standardized Support for Spatial Vector Data. R J 10(1):439-446. https://doi.org/10.32614/rj-2018-009

Penrose R (1955) A generalized inverse for matrices. Math Proc Cambridge. 51(3):406–413. https://doi.org/10.1017/s0305004100030401

Percival J, Tsutsumida N (2017) Geographically Weighted Partial Correlation for Spatial Analysis. GI_Forum. 1:36–43. https://doi.org/10.1553/giscience2017_01_s36

QGIS Development Team (2021) QGIS Geographic Information System. Open Source Geospatial Foundation. http://qgis.org. Accessed 22 February 2021

R Core Team (2020) R. A Language and Environment for Statistical Computing. Vienna, Austria: R Foundation for Statistical Computing.: https://www.r-project.org/. Accessed on 4 December 2020

Schafer J, Opgen-Rhein R, Zuber V, Ahdesmaki M, Silva A P D, Strimmer K (2017) corpcor: Efficient Estimation of Covariance and (Partial) Correlation. R package version 1.6.9. https://CRAN.R-project.org/package=corpcor. Accessed on 4 December 2020

Sievert C (2020) Interactive Web-Based Data Visualization with R, plotly, and shiny. Chapman and Hall/CRC, Florida.

Thomas J J, Cook K A (2006) A visual analytics agenda. IEEE Computer Graphics and Applications. 26(1):10–13. https://doi.org/10.1109/mcg.2006.5

Unwin A, Unwin D (1998) Exploratory Spatial Data Analysis with Local Statistics. J Royal Statistical Soc Ser D Statistician, 47(3), 415–421.

Yu D, Wei Y D, Wu C (2007) Modeling Spatial Dimensions of Housing Prices in Milwaukee, WI. Environ Plan B Plan Des. 34(6):1085–1102. https://doi.org/10.1068/b32119